\documentclass[12pt,preprint]{aastex}
\begin{document}
\title{AN OBSERVATIONAL DETECTION OF THE BRIDGE EFFECT OF VOID FILAMENTS}
\author{Junsup Shim\altaffilmark{1}, Jounghun Lee\altaffilmark{1}, 
Fiona Hoyle\altaffilmark{2}}
\altaffiltext{1}{Astronomy Program, Department of Physics and Astronomy, FPRD, 
Seoul National University, Seoul 151-747, Korea \email{jsshim@astro.snu.ac.kr, jounghun@astro.snu.ac.kr}}
\altaffiltext{2}{Pontifica Universidad Catolica de Ecuador, 12 de Octubre 1076 y Roca, Quito, Ecuador}
\begin{abstract}
{\it The bridge effect of void filaments} is a phrase coined by \citet{PL09b} to explain the correlations found 
in a numerical experiment between the luminosity of the void galaxies and the degree of the straightness of 
their host filaments.  Their numerical finding implies that a straight void filament provides a narrow channel 
for the efficient transportation of gas and matter particles from the surroundings into the void galaxies.  
Analyzing the Sloan void catalog constructed by \citet{pan-etal12}, we identify the filamentary structures in 
void regions and determine the specific size of each void filament as a measure of its straightness.  To avoid 
possible spurious signals caused by the Malmquist bias, we consider only those void filaments whose redshifts are 
in the range of $0\le z\le 0.02$ and find a clear tendency that the void galaxies located in the more straight 
filaments are on average more luminous, which is in qualitative agreement with the numerical prediction.  
It is also shown that the strength of correlation increases with the number of the member galaxies of the void 
filaments, which can be physically understood on the grounds that the more stretched filaments can connect the 
dense surroundings even to the galaxies located deep in the central parts of the voids.  This observational 
evidence may provide a key clue to the puzzling issue of why the void galaxies have higher specific star formation 
rates and bluer colors than their wall counterparts.
\end{abstract}
\section{INTRODUCTION}

Despite its extreme low-density,  a void exhibits a dilute miniature of the cosmic web that interconnects the 
void galaxies \citep[e.g.,][]{sahni-etal94,gottlober-etal03,kreckel-etal11,AS13,alpaslan-etal14}. 
The anisotropic spatial correlation of the tidal shear field is believed to be mainly responsible for the 
formation of the mini-web in a void just as it is for the cosmic web in the whole universe 
\citep{WK93,web96}.  As the tidal shear field develops nonlinear correlations during the evolutionary 
process, the mini-web in a void region should become more and more filamentary. 

An intriguing issue is what effect the filamentary mini-web has on the evolution of the void galaxies and how 
strong it is. This can be addressed by looking for correlations between the intrinsic properties of the void 
galaxies and the characteristics of the void filaments. In fact, the voids should be an optimal ground for the 
investigation of the effect of the filamentary web on the galaxy evolution \citep[e.g., see][]{kreckel-etal11}. 
First of all, since the void filaments are much less intricate than their wall counterparts, it is less difficult to 
identify and characterize them.  
Furthermore, since the densities of the voids are all constrained to extremely low values, the effect of the 
environmental density on the properties of the void galaxies can be controlled and thus it should be easier 
to single out the dependence of the galaxy properties on the filamentary web for the case of the void galaxies.

The mini-filaments that pass through a void are expected to bridge the void galaxies with the surrounding 
denser regions.  Hence, the gas and dark matter from the surroundings can be transported into the void 
regions along the mini-filaments, which would enhance the growth of the void galaxies
 \citep{PL09b,kreckel-etal11}.  
The recent observation of \citet{beygu-etal13} which reported a detection of real-time star-forming activities in the 
void galaxies embedded in the filamentary channels full of HI cold gases is in line with this picture that the 
void galaxies can rapidly grow through the filamentary connection with the surroundings.

The bridge effect of void filaments was first noted and quantitatively investigated in the numerical work of 
\citet{PL09b}  using the Millennium Run semi-analytic galaxy catalog \citep{springel-etal05}. What they found was 
that the intrinsic properties of the void galaxies such as the total mass, luminosity and 
blackhole mass are strongly correlated with the degree of the straightness of the host filaments. 
\citet{PL09b} suggested that the presence of this correlation should be attributed to the dependence of the 
efficacy of gas transportation on the geometrical shapes of the void filaments: the more straight void 
filaments should more efficiently carry gas and matter onto the void galaxies. 
Here, our goal is to detect the bridge effect of the void filament from observational data. 

The upcoming sections are outlined as follows.
In Section \ref{sec:voidfil} we describe how the void filaments are identified from the Sloan void catalog and 
how the degrees of their straightness are measured. In Section \ref{sec:bridge} we report an observational 
detection of the correlation signals between the luminosities of the void galaxies and the degrees of the 
straightness of their host filaments. In Section \ref{sec:con} we summarise the results and discuss 
what clues we can obtain to the formation and evolution of the void galaxies from this observational 
detection as well as what future works would improve the current analysis. Throughout this paper, we assume 
a flat $\Lambda$CDM universe where the cosmological constant $\Lambda$ and cold dark matter (CDM) 
dominate at the present epoch whose initial conditions are described by $\Omega_{m}=0.27,\ 
\Omega_{\Lambda}=0.73$,($h=0.7, \sigma_{8}=0.8$), to be consistent with \citet{pan-etal12}.

\section{IDENTIFYING THE MINI-FILAMENTS FROM THE SLOAN VOIDS}\label{sec:voidfil}

Since there is no unique way to define both the filaments and the voids, it is first necessary to 
decide on which algorithm to use for the identification of both.  In the current analysis, to be consistent 
with \citet{PL09b}, we choose the filament-finding and the void-finding algorithms developed by 
\citet{colberg07} and \citet[][hereafter HV02]{HV02}, respectively. The former is based on the 
minimal spanning tree (MST) technique to find the networking pattern (i.e., MST) in the spatial distributions 
of the point sources and identify the filamentary structures as the most conspicuous cylindrical structures 
in the interconnected MSTs \citep{barrow-etal85,colberg07}. 

In the HV02 algorithm, a void is identified as an underdense region containing only field but no wall galaxies. 
The distance to the fourth nearest galaxy was used as a criterion for the classification of a galaxy: 
For the wall (field) galaxy, $d\le d_{\rm th}$ ($d> d_{\rm th}$) where $d_{\rm th}$ is a given threshold distance 
whose value depends on the parent sample of the galaxies \citep{EP97}. 
The boundary of a void is approximated by the continuous surfaces of the superimposed spheres that fit the 
underdense region \citep{EP97,HV02}. 

\citet{pan-etal12} constructed a catalog of the local voids by applying the HV02 algorithm to the seventh 
Data Release of the Sloan Digital Sky Survey (SDSS DR7)\citep{sdssdr7}. The void catalog contains a total 
of $1055$ giant voids in the redshift range of $0\le z\le 0.107$. It is basically a magnitude-limited sample of the void 
galaxies for each of which information on redshift, equatorial coordinates and the $r$-band magnitude is all available.  
The effective radii ($R_{\rm eff}$) and the numbers of the member galaxies ($N_{\rm g}$) of the voids were found 
to be in the ranges of $9.85 \le R_{\rm eff}/(h^{-1}{\rm Mpc})\le 33.92$ and $2\leq N_{\rm g}\leq 2984$, 
respectively \citep{pan-etal12}. 

As \citet{PL09b} did, we exclude those voids which have less than thirty member galaxies from the void 
catalog of Pan et al. (2012) on the grounds that the filamentary structures are hard to find from those voids. 
A sample of $831$ giant local voids is constructed to which we apply the MST-based filament finding algorithm 
as follows.  For each void in the sample, the member galaxies are regarded as points. The link between 
a randomly chosen point and its nearest point forms an initial MST, which grows and become updated as the other 
points in the same void are sequentially added to it in an decreasing order of the separation distances via an 
iterative search for the nearest points. 

A full MST comes out from each void if all of the points become connected. 
The void filaments are obtained by reducing down a full MST, which requires us to specify  
two parameters: the pruning level $p$ and the separation threshold $d_{c}$.  After the minor branches 
consisting of less than $p$ points are all pruned away, the remaining part of the MST is separated into 
several mini-filaments by disjoining the galaxies whose separation distances are larger than a given threshold 
$d_{c}$ \citep{colberg07}.  For the detailed description of the MST-based filament identification procedures applied 
to the void regions, see \citet{PL09a} and \citet{PL09b}.

As done in \citet{colberg07} and \citet{PL09b}, the best pruning level $p$ is determined as the value at which the 
size distribution of the void filaments becomes stabilized in the large size section, where the size $S$ of a filament is 
defined as the spatial extent of the three dimensional positions of its member galaxies. 
The size distribution of the void filaments, $dN/d\log S$, is calculated as the number of the void filaments 
whose sizes $S$ belong to the differential interval $[\log S, \log S+d\log S]$ \citep{PL09a}.

\begin{figure}[ht]
\begin{center}
\plotone{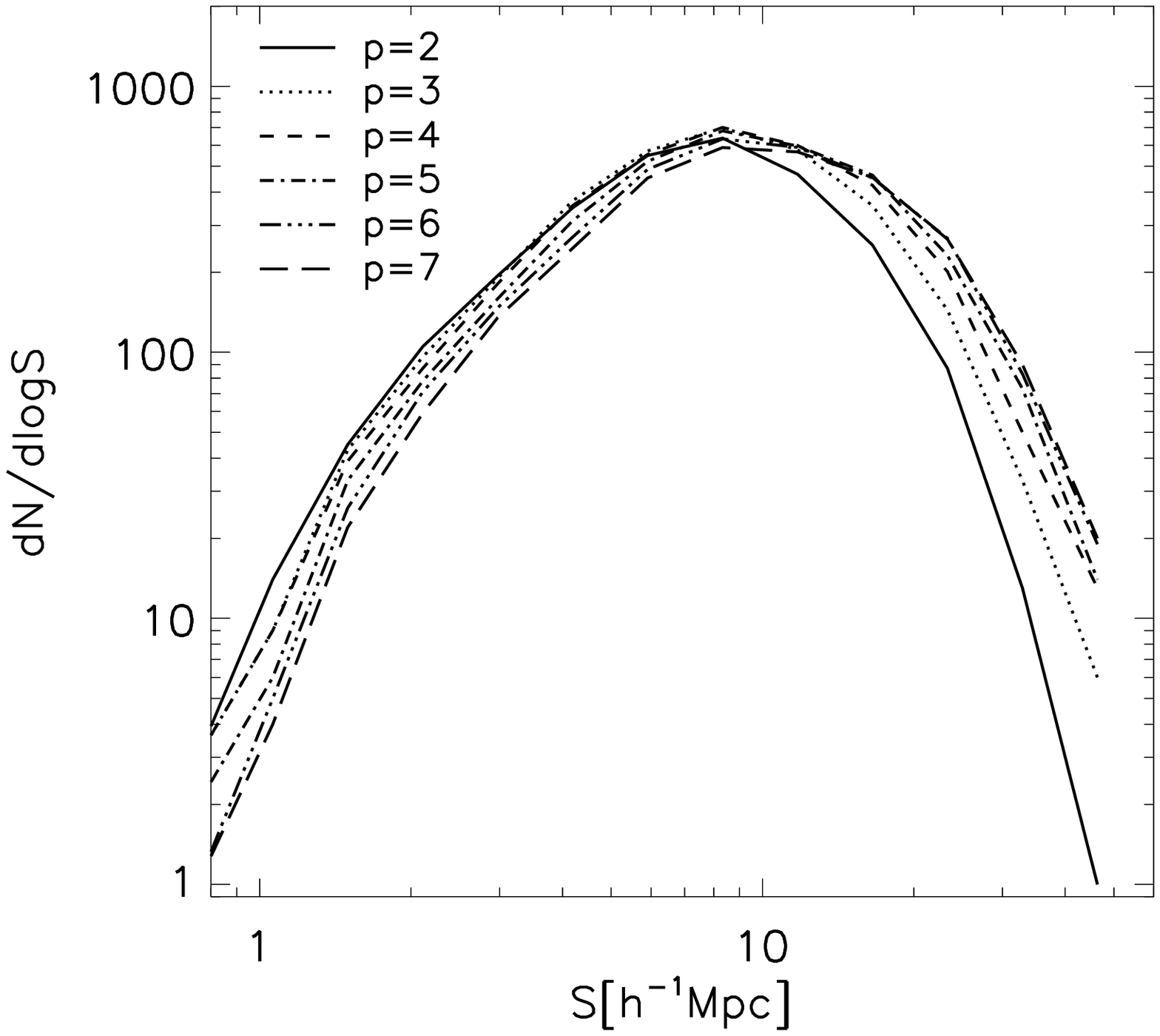}
\caption{Size distribution of the void mini-filaments for six different cases of the pruning level $p$. 
Here, a void filament represents the primary structure of a MST from which the minor twigs having  less than 
$p$ nodes are removed and its size $S$ is define as its spatial extent. It is clear that the convergence of 
$dN/d\log S$ in the large $S$-section appears at $p\ge 5$.}
\label{fig:dndS}
\end{center}
\end{figure}

Varying the pruning level $p$ from $2$ to $7$, we repeatedly apply the MST-based filament finder to 
the $831$ giant voids and determine $dN/d\log S$.  
Figure \ref{fig:dndS} plots the size distributions of the void filaments from our void sample, 
$dN/d\log S$, for six different cases of the pruning level $p$.  For the case of $p\le 4$, the functional form 
of $dN/d\log S$ fluctuates with the change of $p$ in the large size section ($S \ge 10\,h^{-1}$Mpc). However, 
note that it becomes stabilized against the change of $p$ if $p$ reaches and exceeds $5$, which leads us to 
determine the best pruning level to be $p=5$. It is interesting to recall that in the numerical experiment of \citet{PL09b} 
the size distribution of the void filaments from the Millennium semi-analytic galaxy catalog was shown to become 
stabilized at the same pruning level $p=5$.

\begin{figure}[ht]
\begin{center}
\plotone{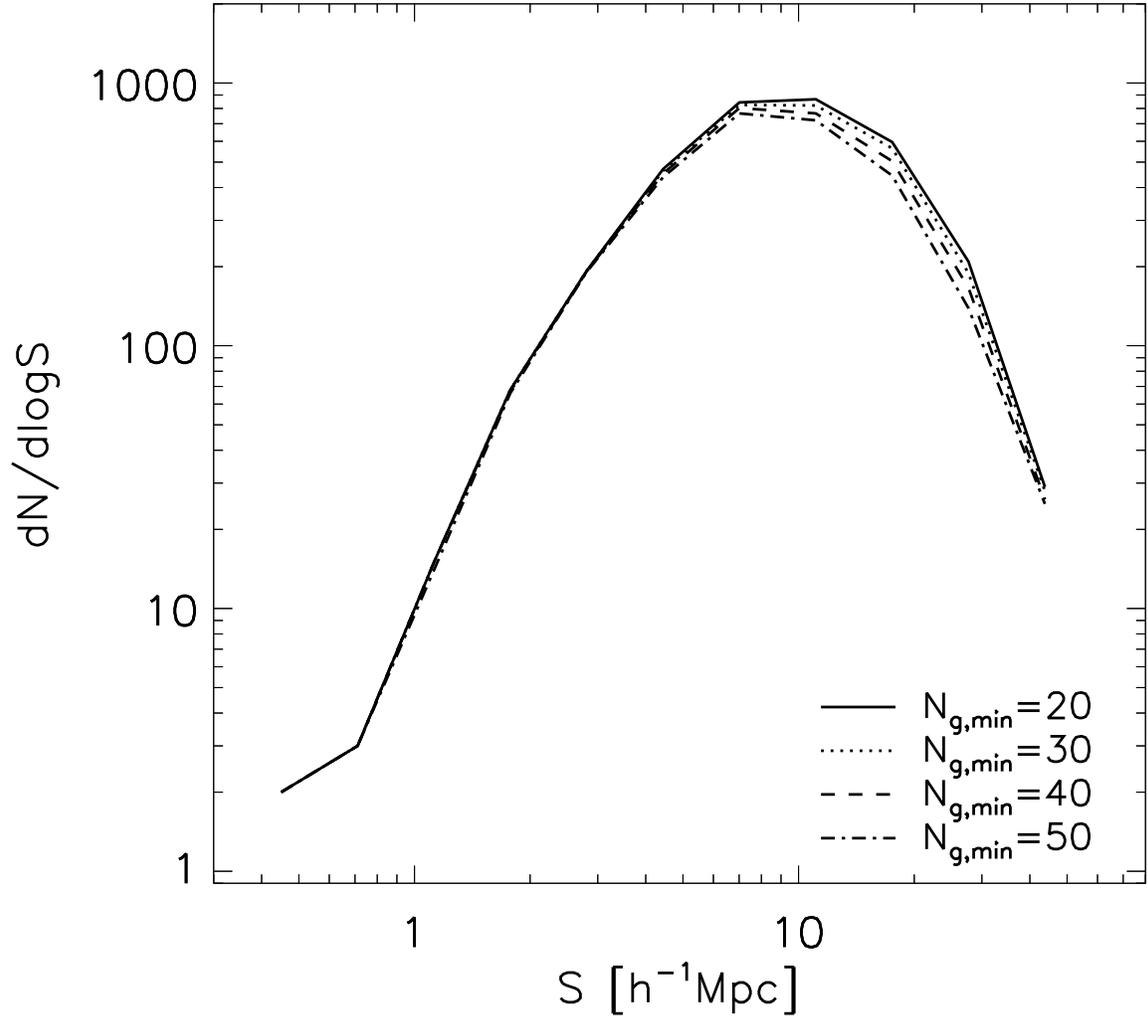}
\caption{Number distribution of the sizes of void filaments in interval $[\log S,\log S+d\log S]$ 
for four various minimum number of galaxies in a void $N_{g,min}$ applied for the giant void selection.} 
\label{fig:vary_dndS}
\end{center}
\end{figure}

Since the best value of $p$ depends on the number of the sample galaxies, we investigate how the best  
value of $p$ changes if the minimum number of the void galaxies, say $N_{\rm g,min}$, differs from the threshold value 
of $30$ adopted by \citet{PL09b}.  Varying the value of $N_{\rm g,min}$ from $20$ to $50$, we repeat the whole process to 
evaluate $dN/d\log S$, which is shown in Figure \ref{fig:vary_dndS}. Although the number density of the void 
filaments with large sizes ($S\ge 5\,h^{-1}$Mpc) decreases substantially as the value of $N_{\rm g, min}$ increases, 
it is found that the four different cases of $N_{\rm g, min}$ yield the same pruning level $p=5$.  
Furthermore, the correlation strength between the luminosity and the straightness of the void filaments quantified 
by the Pearson product moment correlation coefficient $r$ turns out to be robust against the change of $N_{\rm g,min}$ 
(see section \ref{sec:bridge}). Throughout this paper, we set $N_{\rm g,min}$ at $30$ to be consistent with \citet{PL09b}. 

To determine the value of $d_{c}$, \citet{PL09b} sought for the value which maximizes the number of the 
void filaments consisting of four and more member galaxies. They excluded the filaments with less than 
four member galaxies since those filaments should always have high degree of straightness. 
\citet{PL09b} noted that the number of the void filaments with four and more member galaxies 
varies strongly with $d_{c}$ and reached the highest value when $d_{c}$ is equal to $\bar{d}+\sigma_{d}$ 
where $\bar{d}$ and $\sigma_{d}$ are the mean and the standard deviation of the separation distances of the 
member galaxies belonging to the unpruned MSTs, respectively. Following \citet{PL09b}, we calculate  
the values of $\bar{d}$ and $\sigma$ from the unpruned MSTs of the void galaxies found in the 
Sloan void catalog of \citet{pan-etal12},  we determine the critical separation distance $d_{c}$ as $d+\sigma_{d}$.

\begin{figure}[ht]
\begin{center}
\plotone{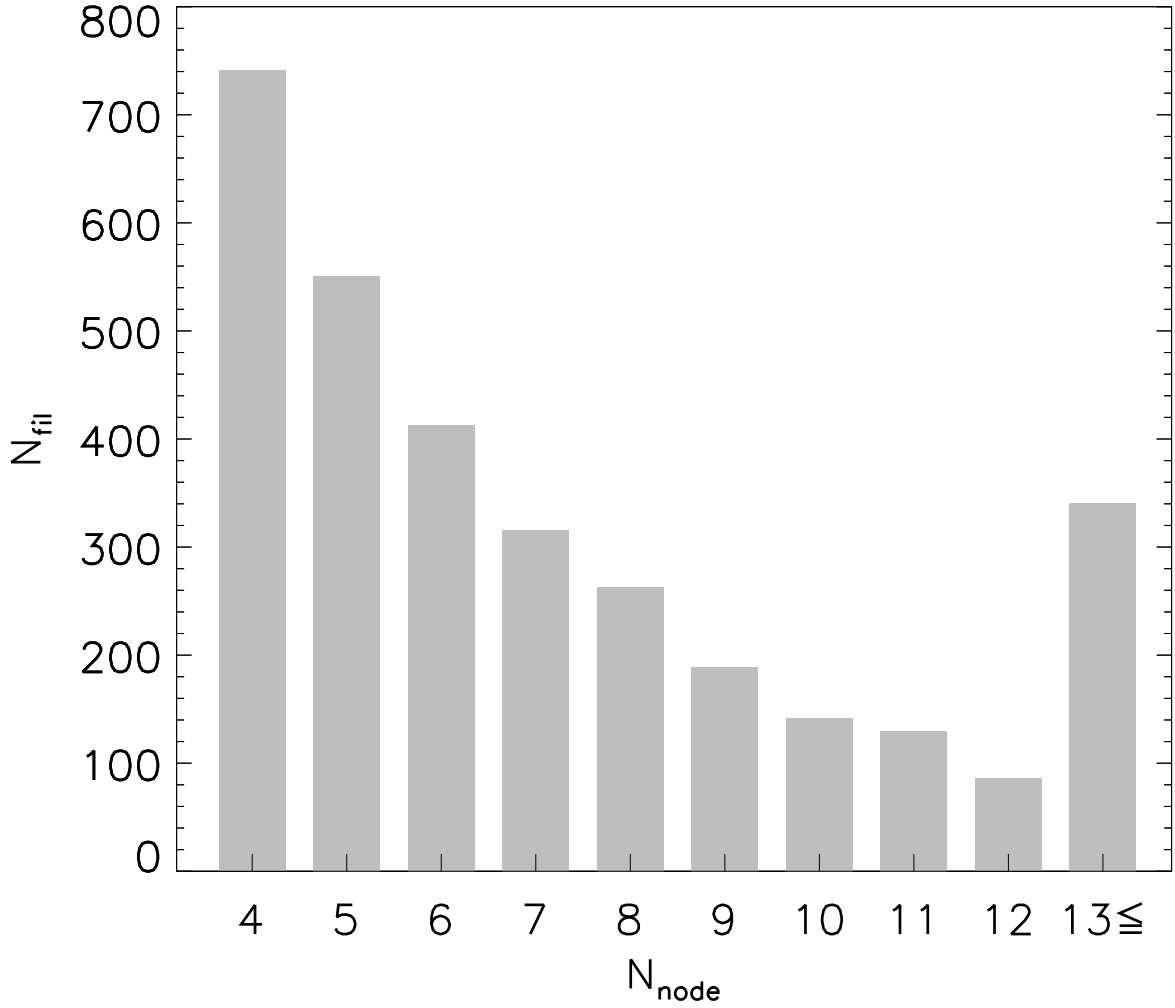}
\caption{Number counts of the void filaments having four or more nodes from the Sloan void catalog 
by \citet{pan-etal12} as a function of the node number, $N_{\rm node}$.}
\label{fig:n_dis}
\end{center}
\end{figure}

Among a total of $3172$ void filaments found in the redshift range of $0\le z\le 0.107$ from the Sloan void catalog 
of \citet{pan-etal12}, the richest one turns out to have $38$ nodes where the node is a jargon of the MST algorithm 
referring to a member galaxy of a filament \citep{colberg07}. 
Figure \ref{fig:n_dis} shows the number of the void filaments $N_{\rm fil}$ as a function of the number of  their 
nodes, $N_{\rm node}$, in our sample. For the right-most bin, we plot the accumulated numbers of the void 
filaments which have $13$ or more nodes. 
As can be seen, the number of the void filaments $N_{\rm fil}$ decreases sharply with the number of the 
nodes $N_{\rm node}$, indicating that the filaments are relatively short in the voids compared with the 
wall counterparts.

\section{CORRELATIONS BETWEEN THE GALAXY LUMINOSITY AND THE SPECIFIC SIZES 
OF VOID FILAMENTS}\label{sec:bridge}

In the original work of \citet{PL09b} based on a N-body simulation, the linearity, $R_{\rm L}$, of each 
filament, defined as the ratio of its end-to-end separation to its total length was used as an indicator of 
the straightness where the total length of a void filament was defined as the sum of the separation distances 
between the adjacent nodes \citep[see also][]{colberg07,PL09a}. 
However, in our current work dealing with the void filaments from real observations identified in redshift 
space,  the linearity $R_{\rm L}$ may not be a good indicator of the degree of the straightness of a void 
filament since the total length of a void filament that enters in the definition of $R_{\rm L}$ depends 
sensitively on how accurately the positions of the void galaxies are measured.

The systematic errors associated with the redshift-space measurement of $R_{\rm L}$  may severely 
contaminate the measurement of the correlations between the luminosity of the void galaxies and the 
degree of the straightness of the hosting void filaments. Therefore, instead of $R_{\rm L}$, we use the 
specific size of each void filament, $\tilde{S}$, defined as the size per node as an indicator of the straightness. 
The concept of the specific size of a filamentary structure was first introduced by \citet{SL13} who showed that 
the filamentary structures having larger specific sizes tend to have more straight shapes 
\citep[see Figure 6 in][]{SL13}.

For each void filament in our sample, we also determine the mean absolute value , $\vert\bar{M}_{r}\vert$, of the absolute 
$r$-band magnitudes of the member galaxies. Since the absolute magnitude $M_{r}$ of a galaxy has a negative value and 
what matters in the current analysis is not the sign but the magnitude of $M_{r}$, we take the absolute value of 
$\vert\bar{M}_{r}\vert$ to investigate the correlation with the specific size of the hosting void filament. From there on, 
$\vert\bar{M}_{r}\vert$ denotes the mean absolute value of $M_{r}$ averaged over the member galaxies of each void 
filament in the sample.

\begin{figure}[ht]
\begin{center}
\plotone{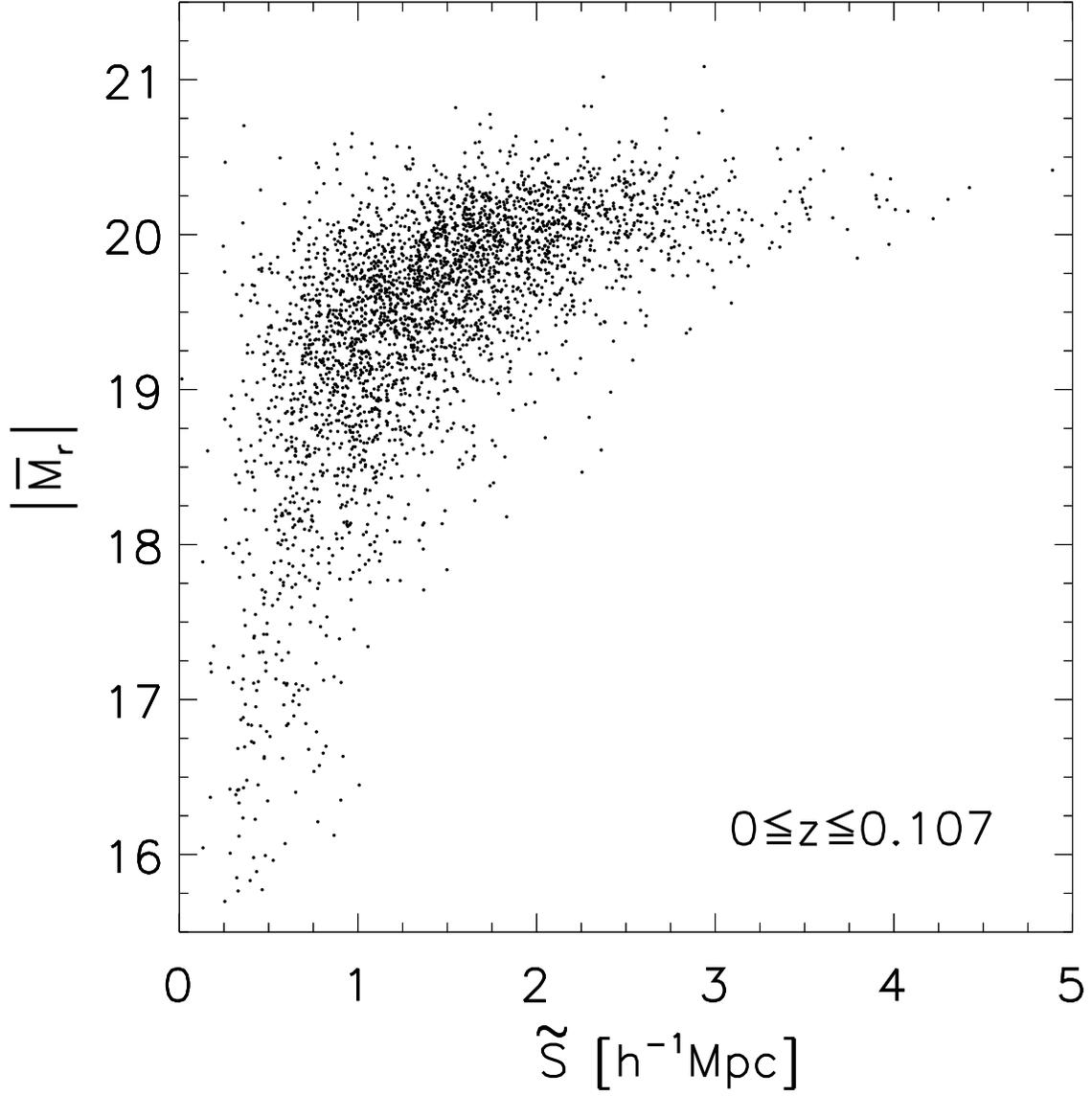}
\caption{Specific sizes of the $3172$ void filament $\tilde{S}\equiv S/N_{\rm node}$ (size per node) 
versus the mean absolute $r$-band magnitude of its member galaxies $\vert\bar{M}_{r}\vert$ from our sample.}
\label{fig:scatter}
\end{center}
\end{figure}

Figure \ref{fig:scatter} plots how the void filaments in our sample are scattered in the plane spanned by 
$\vert\bar{M}_{r}\vert$  and $\tilde{S}$, showing an obvious trend that the averaged luminosity of the void galaxies
increases with the degree of the straightness of their host filament.  Before interpreting this trend as an observational
evidence for the bridge effect of the void filaments, it should be critical to examine the possibility that the trend 
might be just a consequence of the Malmquist bias \citep{malmquist36}.  Given that the Sloan void catalog of 
\citet{pan-etal12} is a magnitude-limited sample,  the high-$z$ void filaments must be  biased being composed 
only of very luminous galaxies. Since the mean separation distance obtained from a sample including only the very luminous 
galaxies should be larger than than from a sample including the dim galaxies, the separation threshold $d_{c}$ that is used as a 
criterion to separate each MST into filaments (see Section \ref{sec:voidfil}) tends to have a larger value for the case of the 
high-$z$ voids. Therefore, the void filaments identified at high-redshifts via the MST algorithm are likely to have more extended 
spatial sizes per node, which would in turn create a spurious signal of correlation between $\vert\bar{M}_{r}\vert$ and $\tilde{S}$. 

There are two different ways to remove the Malmquist bias. The first way is to construct a volume-limited 
sample of the void galaxies up to $z=0.107$ by excluding the dim galaxies below the corresponding threshold magnitude limit. 
The second one is to consider only the nearby void filaments located at the lowest redshifts. We choose the second direction for 
the following reason. As mentioned in \citet{PL09b}, the bridge effect is believed to be related to the effect of the anisotropic tidal 
shear field on the evolution of the void galaxies (see also Section \ref{sec:con} for a full discussion). Thus, to detect a genuine 
signal of the bridge effect, it is important to identify the void filaments whose straightness can reflect well the anisotropic nature of 
the tidal shear field. If the void filaments are found in the volume-limited sample of the luminous galaxies above the corresponding 
magnitude limit, it is likely that the degree of their straightness fails to represent the degree of anisotropy of the tidal shear field 
because the luminous galaxies are the biased tracers of the matter density field. As the luminous galaxies are expected to have 
formed at the high peaks of the matter density field whose gravitational collapse proceeds quite isotropically 
\citep{bbks86,bernardeau94}, the void filaments identified from the volume-limited sample would fail to capture the unbiased 
features of the tidal shear field, no matter how straight they are. In other words, to represent the anisotropic nature of the tidal 
shear field and its effect on the evolution of the filament galaxies in void regions, the void filament must be identified from the 
unbiased tracers including the dim galaxies (or low-mass halos). That is why we choose to use a magnitude-limited low-z sample 
instead of a volume limited high-z sample, even though the latter is a homogeneous one. 

\begin{figure}[ht]
\begin{center}
\plotone{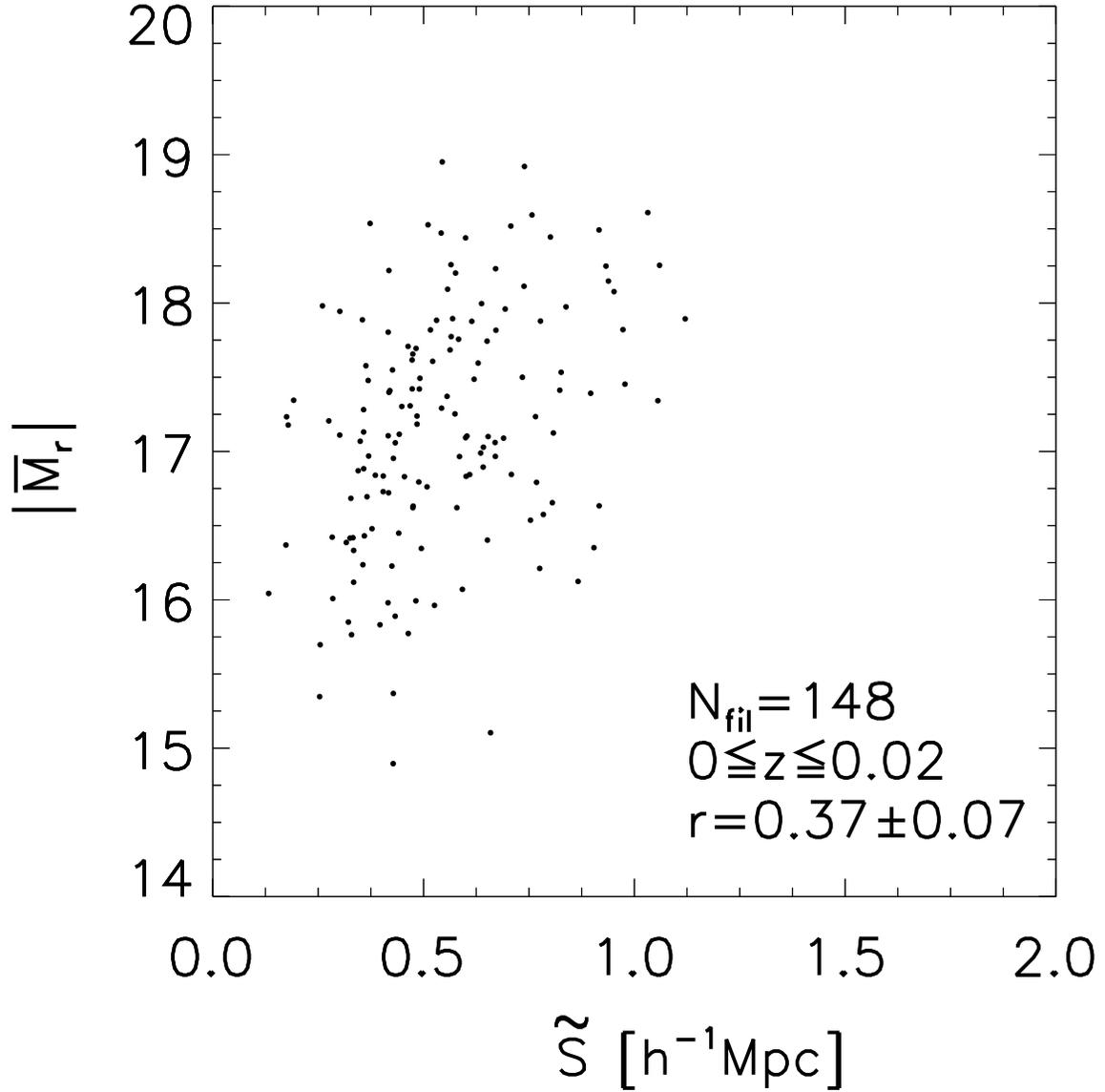}
\caption{A total of $148$ void filaments with four or more member galaxies at $0\le z\le 0.02$ from the Sloan 
void catalog of \citet{pan-etal12} in the $\tilde{S}-\vert\bar{M}_{r}\vert$ plane.}
\label{fig:main_scatter}
\end{center}
\end{figure}

A magnitude-limited sample of the nearby void filaments whose redshift is low enough to avoid any false signal caused by the 
Malmquist bias and whose size is large enough to produce a significant signal is found to correspond to the redshift range 
$0\le z\le 0.02$. We find a total of $148$ void filaments whose relation between $\tilde{S}$ and $\vert\bar{M}_{r}\vert$ is shown 
in Figure \ref{fig:main_scatter}. As can be seen, even at this low redshift bin $0\le z\le 0.02$, the absolute mean values, 
$\vert\bar{M}_{r}\vert$, of the void filaments still appear to be correlated with their specific sizes $\tilde{S}$. To quantify the 
correlation between $\vert\bar{M}_{r}\vert$ and $\tilde{S}$ of the $148$ void filaments, we calculate the Pearson product 
moment coefficient, $r$, as \citep{WJ12}:
\begin{equation}
\label{eqn:r}
r = \frac{\langle\Delta\vert\bar{M}_{r}\vert\,\Delta{\tilde{S}}\rangle}{\left[\langle(\Delta \vert\bar{M}_{r}\vert)^{2}\rangle
\langle(\Delta\tilde{S})^{2}\rangle\right]^{1/2}} \, ,
\end{equation}
where $\Delta\vert\bar{M}_{r}\vert\equiv \vert\bar{M}_{r}\vert-\langle\vert\bar{M}_{r}\vert\rangle$ and 
$\Delta\tilde{S}\equiv \tilde{S}-\langle\tilde{S}\rangle$. 
Here $\langle\vert\bar{M}_{r}\vert\rangle$ and $\langle\tilde{S}\rangle$ are the ensemble averages of the mean 
absolute $r$-band magnitudes and the specific sizes over the $148$ void filaments, respectively, and 
$\langle(\Delta\vert\bar{M}_{r}\vert)^{2}\rangle$ and $\langle(\Delta\tilde{S})^{2}\rangle$ are the corresponding 
rms fluctuations, respectively.

From the $148$ void filaments with four or more member galaxies located at $0\le z\le 0.02$, we find $r=0.37\pm0.07$ 
where the errors associated with the estimate of $r$ is also calculated as \citep{WJ12}
\begin{equation}
\label{eqn:sigma_r}
\sigma_{r}=\frac{1-r^{2}}{(N_{\rm fil}-1)^{1/2}}\, ,
\end{equation}
where $N_{\rm fil}$ is the number of the void filaments.
To see at what confidence level the null hypothesis of no correlation between $\vert\bar{M}_{r}\vert$ and $\tilde{S}$ is 
rejected, we perform the student-$t$ statistics as \citep{WJ12}
\begin{equation}
\label{eqn:student}
t=\frac{r{(N_{\rm fil}-2)}^{1/2}}{(1-r^{2})^{1/2}} \, ,
\end{equation}
where $N_{\rm fil}-2$ is the degree of freedom of the $t$-variable.  Plugging $r=0.37$ and $N_{\rm fil}=148$ 
into Equation (\ref{eqn:student}), we find that the null hypothesis is rejected at $99.999\%$ confidence level. 

\begin{figure}[ht]
\begin{center}
\plotone{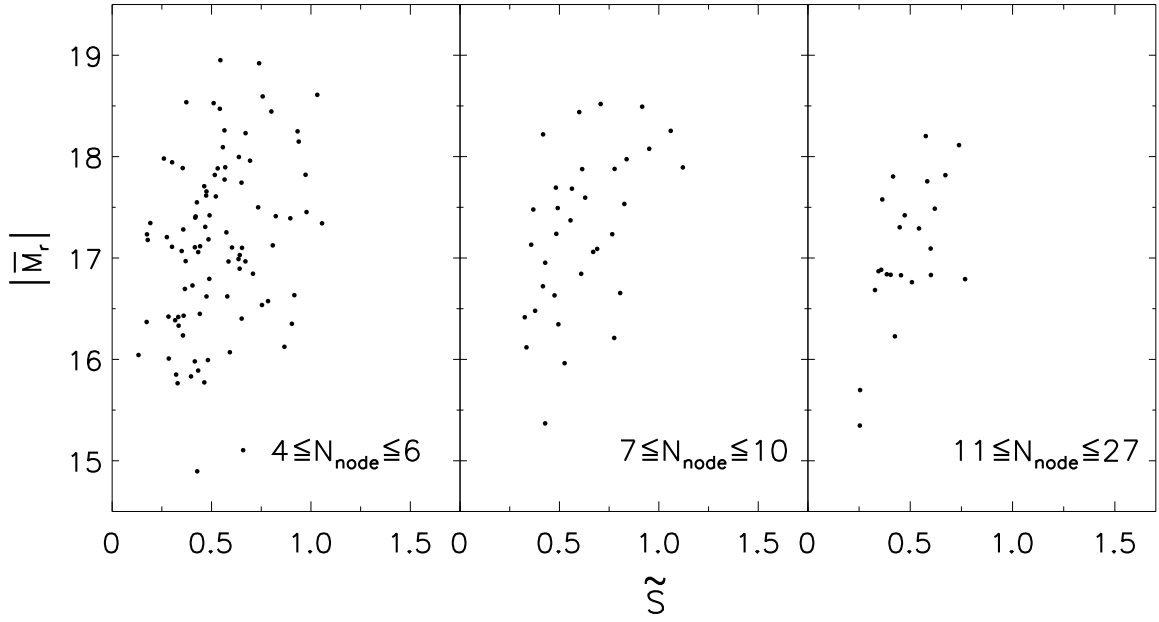}
\caption{Same as Figure \ref{fig:main_scatter} but for three different bins of the node number $N_{\rm node}$.}
\label{fig:3panel_scatter}
\end{center}
\end{figure}

Now that we find an observational evidence for the correlation between $\tilde{S}$ and $\vert\bar{M}_{r}\vert$, 
we would like to investigate if the strength of correlation depends on the richness of the void filaments (i.e., the 
number of nodes, $N_{\rm node}$). Dividing the void filaments at $0\le z\le 0.02$ into three $N_{\rm node}$-bins, 
we calculate $r$ separately by using only those void filaments belonging to each $N_{\rm node}$-bin. 
Figure \ref{fig:3panel_scatter} shows the  $\tilde{S}$-$\vert\bar{M}_{r}\vert$ scatter plots for three different bins of 
$N_{\rm node}$.  Table \ref{tab:result} also lists the mean redshifts, the numbers of the void filaments and the 
estimated values of the Pearson product moment correlation coefficient $r$ between $\tilde{S}$ and $\vert\bar{M}_{r}\vert$ 
(the fourth column) for three $N_{\rm node}$-bins. 
 As can be read, the richer filaments exhibit stronger correlations between $\vert\bar{M}_{r}\vert$ and $\tilde{S}$.  The value 
of $r$ reaches as high as $0.6$ for the richest case of $11\le N_{\rm node}\le 27$. It is also worth noting that the increase 
of $r$ with $N_{\rm node}$ is not caused by the redshift-difference among the three $N_{\rm node}$-bins.

\begin{figure}[ht]
\begin{center}
\plotone{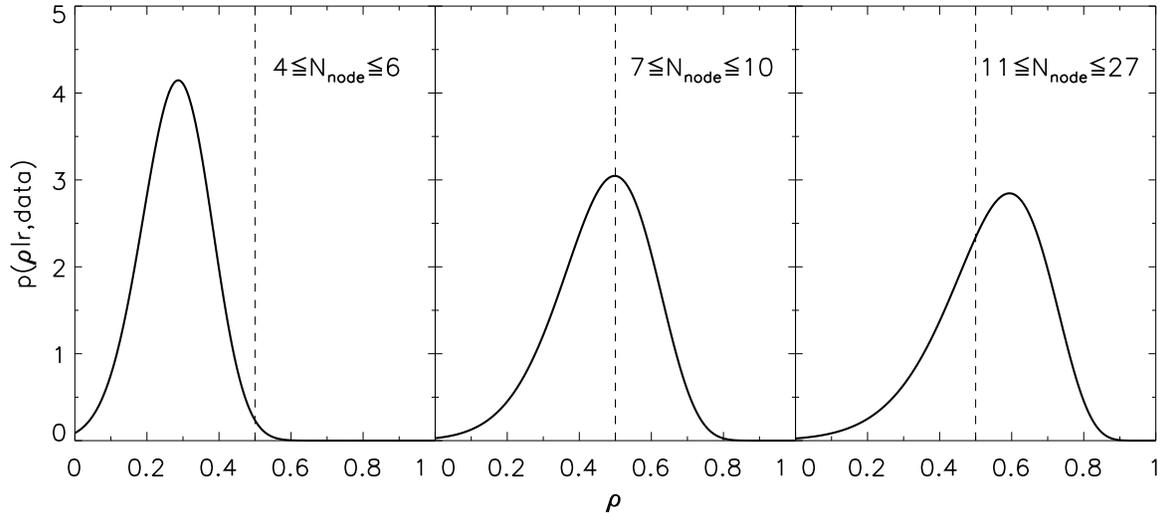}
\caption{Posterior probability density distribution of the real correlation coefficient $\rho$ provided that the Pearson 
product moment correlation coefficient $r$ has the resulting value from our sample at $0\le z\le 0.02$ for three different 
cases of $N_{\rm node}$ (solid line). In each panel the dashed line corresponds to the case of a mild correlation 
($\rho = 0.5$).}
\label{fig:bayes}
\end{center}
\end{figure}

We have so far followed the frequentist's approach to quantify the correlations between $\vert\bar{M}_{r}\vert$ and 
$\tilde{S}$. However, the reliability of the frequentist's approach depends on the sample size. Given that the 
numbers of the low-$z$ void filaments at each $N_{\rm node}$-bin are not large enough to 
guarantee the success of the frequentist's approach, we also take the Bayesian approach by computing 
the posterior probability density distribution of the true correlation coefficient, $\rho$ as \citep{WJ12}
\begin{equation}
p(\rho|r,{\rm data}) \propto \frac{(1-{\rho}^{2})^{\frac{N_{\rm fil}-1}{2}}}{(1-r\rho)^{N_{\rm fil}-\frac{3}{2}}}
\left[1+\frac{1}{N_{\rm fil}-\frac{1}{2}}\frac{1+r\rho}{8}+\cdot\cdot\cdot\right]\, ,
\end{equation}
where $p(\rho\vert r,{\rm data})$ represents the conditional probability of $\rho$ provided that the Pearson product 
moment correlation coefficient has the value of $r$ at each $N_{\rm node}$-bin. 
Here, we assume a flat prior for $p(r)$ and normalized $p(\rho|r, {\rm data})$ to satisfy 
$\int p(\rho | r, {\rm data})=1$. 
The left, middle and right panels of Figure \ref{fig:bayes} show $p(\rho| r, {\rm data})$, for the cases of 
$4\le N_{\rm node}\le 6$, $7\le N_{\rm node}\le10$ and $11\le N_{\rm node}\le 27$, respectively. 
As can be seen, the maximum probability is achieved when the true correlation coefficient exceeds $0.6$ 
for the case of the richest filaments, supporting the results obtained from the frequentist's approach.

The observed trend that the correlation between $\vert\bar{M}_{r}\vert$ and $\tilde{S}$ becomes stronger with the 
richness of the void filament can be understood as follows. The richer void filaments correspond to deeper gravitational 
potential well which can accommodate larger amounts of cold gas, which in turn can be transported even to the 
central regions of the voids from the surroundings if the void filaments are more spatially extended. 
In consequence, the mean luminosity of the void galaxies located in the richer filaments would develop 
more sensitive dependence on the spatial extents of the filamentary networking.  

\begin{figure}[ht]
\begin{center}
\plotone{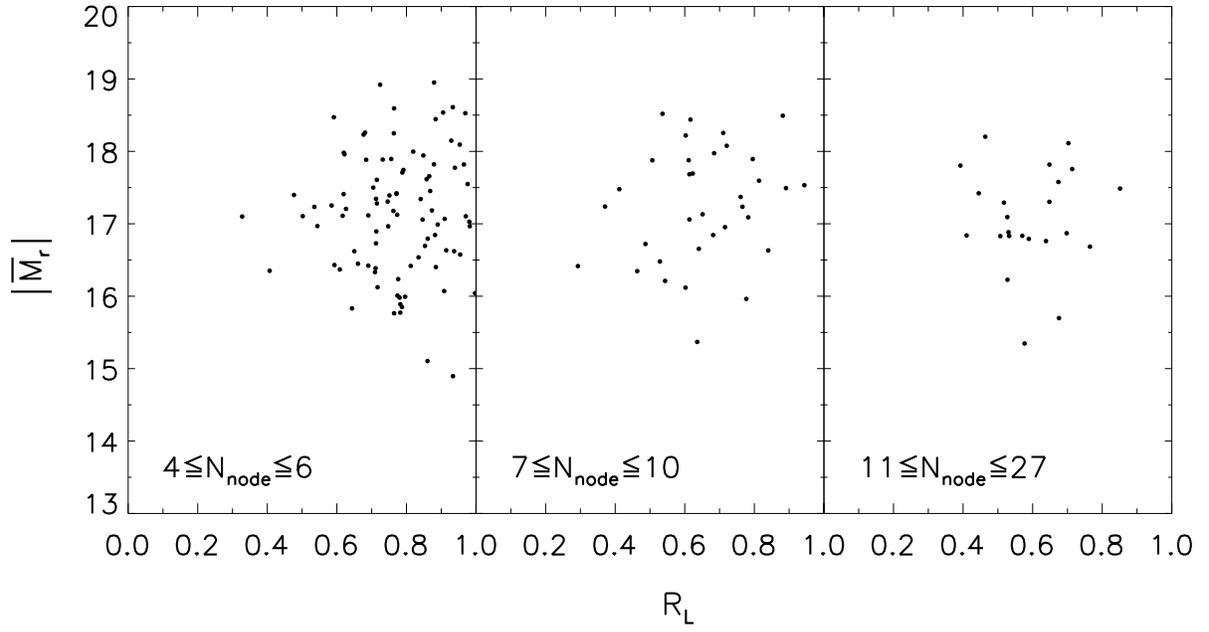}
\caption{Same as Figure \ref{fig:3panel_scatter} but replacing the specific size $\tilde{S}$ by the linearity $R_{\rm L}$ 
to represent the degree of the straightness of the void filaments.}
\label{fig:rl_mr}
\end{center}
\end{figure}

\begin{figure}[ht]
\begin{center}
\plotone{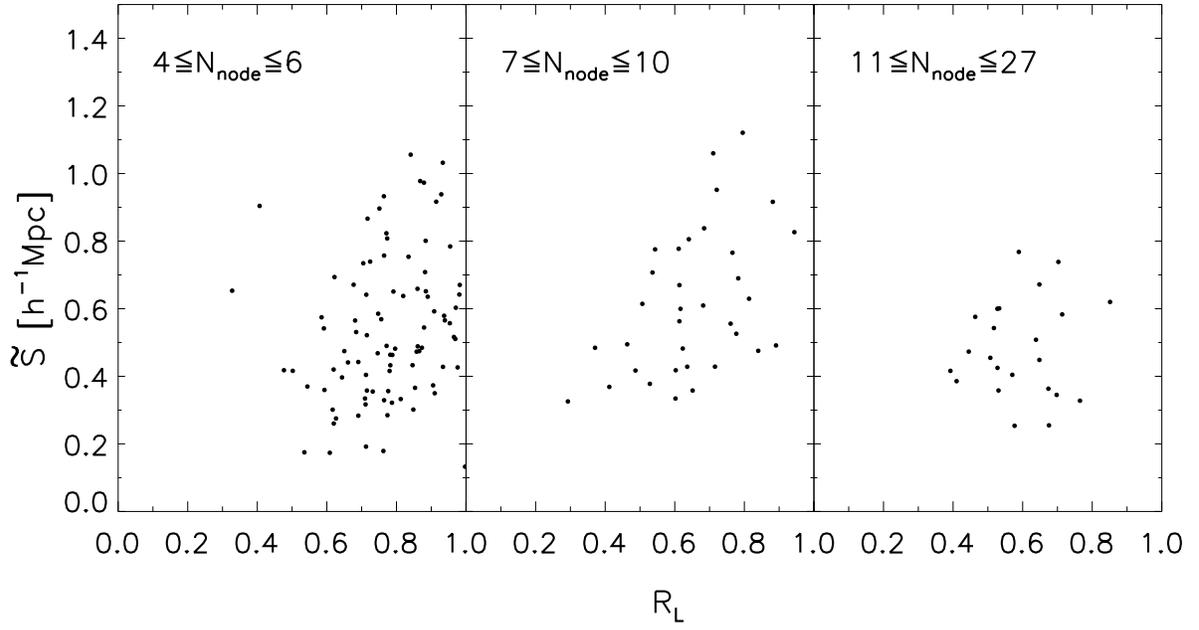}
\caption{Specific sizes $\tilde{S}$ of $148$ void filaments versus their linearities $R_{\rm L}$.}
\label{fig:ts_rl}
\end{center}
\end{figure}

It may be worth examining the dependence of the final result on how to quantify the degree of the straightness of 
the void filaments. As in the original work of \citet{PL09b}, we measure the linearity $R_{\rm L}$ of the $148$ 
void filaments and investigate the correlations between $R_{\rm L}$ and $\vert\bar{M}_{r}\vert$ at three 
$N_{\rm node}$-bins, which are shown in Figure \ref{fig:rl_mr}. The Pearson product moment correlation 
coefficients between $R_{\rm L}$ and $\vert\bar{M}_{r}\vert$ for the three different $N_{\rm node}$-bin are 
listed in the fifth column of Table \ref{tab:result}. As can be seen, no correlation is found between $R_{\rm L}$ and 
$\vert\bar{M}_{r}\vert$, which is inconsistent with the result of \citet{PL09b} whose numerical analysis found the existence 
of a strong correlation $r=0.67$ between them.
To understand where this inconsistency stems from, we examine how strongly $R_{\rm L}$ and $\tilde{S}$ are 
correlated with each other for the three $N_{\rm node}$-bins, the results of which are shown in Figure \ref{fig:ts_rl} and 
listed in the sixth column of Table \ref{tab:result} . As can be seen, a signal of mild correlation is found  
for the case of $7\le N_{\rm node}\le 11$ but no signal at all for the other two cases. 

Note that for the case of $4\le N_{\rm node}\le 6$, the linearity tends to be biased toward high values of 
$0.4\le R_{\rm L}\le 1$. As explained in \citet{PL09b}, for the case that the filaments have small number of nodes, 
the linearity $R_{\rm L}$ is prone to high values, failing to represent well the degree of filament straightness.  
For the case of the rich filaments with large number of nodes, the linearity $R_{\rm L}$ is a good indicator of 
the filament straightness only provided that the positions of the nodes are measured with high accuracy. 
When the measurements are done in the redshift space, however, the redshift-space errors 
would contaminate the measurement of each of the positions of the nodes, which would propagate into 
the errors in the measurements of the total lengths of the filaments. The larger numbers of the nodes 
a filament has, the larger amount of propagated error would contaminate the measurement of its total length 
which would in turn cause large uncertainties in the estimate of its linearity. 

\section{SUMMARY AND DISCUSSION}\label{sec:con}
 
The aim of this work was to observationally test the scenario suggested by \citet{PL09b} that the void galaxies 
located in the more straight filaments would be more luminous since the efficacy of the gas accretion along the void 
filaments into the void galaxies increases with the degree of the straightness of the void filaments. The catalog of 
the giant galaxy voids constructed by \citet{pan-etal12} from the SDSS DR7 datasets with the help of the HV02 
void finding algorithm \citep{HV02} has been used as the parent sample and the MST-based filament finder was 
employed to identify the filamentary structures in the galaxy voids. 

Determining the specific size (size per node, $\tilde{S}$) of each void filament as a 
measure of its  straightness and the mean absolute value of the absolute $r$-band magnitude $\vert\bar{M}_{r}\vert$ of the 
member galaxies, we investigated if and how $\vert\bar{M}_{r}\vert$ is correlated with $\tilde{S}$ on average by measuring the 
Pearson product moment correlation coefficient, $r$, between the two quantities. 
To avoid a false signal of correlation between $\tilde{S}$ and $\vert\bar{M}_{r}\vert$ caused by the Malmquist bias,
we have focused only on the lowest redshift for the detection of the bridge effect. 
From a total of $148$ void filaments consisting of four or more member galaxies (nodes) in the redshift range 
of $0\le z\le 0.02$, it is  found to be $r=0.37\pm 0.07$.  
Although this observational signal is not quantitatively consistent with the numerical estimate, $r=0.67$, by \citet{PL09b} based 
on the Millennium run semi-analytic galaxy catalog \citep{springel-etal05}, it is a five sigma signal and thus may be regarded 
as the observational evidence for the existence of the bridge effect of the void filaments. 

We have also inspected if and how the strength of correlation between $\vert\bar{M}_{r}\vert$ and $\tilde{S}$ 
changes with the node number $N_{\rm node}$ of the void filaments (i.e., the {\it richness} of the void filaments) 
and noted a trend that the value $r$ gradually increases with the richness of the void filament.  Our explanation 
for this trend is as follows. A richer filament must correspond to a deeper well of the gravitational potential that 
could accommodate larger amounts of gas inflow from the surroundings.  In consequence the void galaxies 
located in a richer filament would develop more sensitive dependence on the degree of the straightness of void 
filament. 

The straightness of the void filaments reflects how anisotropic the spatial distributions of the void galaxies 
are. As the voids evolve and their densities decrease, they would become more susceptible to the external 
tidal forces, which in turn augment the degree of the anisotropy of the spatial distributions of the void 
galaxies. In other words, the capacity of the void filaments as a transportation channel would increase in the 
nonlinear stage of the void evolution where the tidal shear field develop nonlinear correlations. 
In consequence, the star formation activity in the void galaxies  would become stimulated by the 
efficient supply of the cold gases along the straight void filaments.  
The presence of the bridge effect of void filaments that has been confirmed by our observational analysis 
may provide an important clue to the mechanism through which the void galaxies become gas richer, bluer 
colored, and have higher specific star formation rates than their wall counterparts 
\citep[e.g.,][]{rojas-etal04,kreckel-etal11,beygu-etal13,ricciardelli-etal14}. 

However, we have yet to answer the fundamental question of what caused the bridge effect of void 
filaments.  A more comprehensive study based on cosmological hydrodynamic simulations will be desirable to 
quantitatively understand why and how the accretion of cold gas occur more efficiently along the straight filaments 
that emerge during the nonlinear stage of the void evolutions. 
Also, in the original work of \citet{PL09b}, not only the mean luminosity but also the mean mass of 
the central blackholes of the void galaxies were found to be higher in the more straight void filaments. 
In fact, it was the mean mass of the central blackholes that yielded the strongest correlation with the 
degree of the straightness of the void filaments.  Although it is very hard to measure directly the masses of the 
central blackholes, the X-ray luminosities of the void galaxies with active galactic nuclei (AGN), if any, 
could be used to quantify how massive their central blackholes are.  
Our future work will focus on searching for the AGN galaxies in the void regions and exploring 
observationally if and how their AGN activities  are correlated with the degree of the straightness of the void 
filaments. 

\acknowledgments

We thank a referee for his/her very enlightening comments on the Malmquist bias.
JS was financially supported by the Basic Science Research Program through the National 
Research Foundation of Korea (NRF) funded by the Ministry of Education (NRF-2014H1A8A1022479).
JL was financially supported by the Basic Science Research Program through the NRF funded by the Ministry 
of Education (NO. 2013004372) and by the research grant from the NRF to the Center for Galaxy 
Evolution Research  (NO. 2010-0027910). FH acknowledges support from Pontifica Universidad Catolica del Ecuador.

\clearpage
\clearpage
\clearpage
\clearpage
\clearpage
\clearpage
\clearpage
\clearpage
\clearpage
\clearpage
\begin{deluxetable}{cccccc}
\tablewidth{0pt}
\setlength{\tabcolsep}{5mm}
\tablecaption{Richness ($N_{\rm node}$), mean redshift ($\bar{z}$), number of the void filaments ($N_{\rm fil}$)
and Pearson product moment correlation coefficients ($r$). }
\tablehead{
$N_{\rm node}$ & $\bar{z}$ & $N_{\rm fil}$ & 
r($\tilde{S},\vert\bar{M}_{r}\vert$) &
r($R_{L},\vert\bar{M}_{r}\vert$) &    
r($R_{L},\tilde{S}$)}   
\startdata
$[4,\ 6]$   & $0.012\pm0.005$ & $91$ & $0.29\pm0.10$ & $0.04\pm0.10$ & $0.21\pm0.10$\\
$[7,\ 10] $ & $0.014\pm0.003$ & $34$ & $0.50\pm0.13$ & $0.21\pm0.16$ & $0.46\pm0.14$\\
$[11,\ 27]$ & $0.013\pm0.004$ & $23$ & $0.60\pm0.13$ & $0.00\pm0.21$ & $0.10\pm0.21$\\
\enddata
\label{tab:result}
\end{deluxetable}
\end{document}